\documentclass{aa}
\usepackage{graphicx}
\usepackage{txfonts}
\def \mdot   {{\hbox{$\skew3\dot M$}}}

\begin{document}
   \title{The ionization structure of early-B supergiant winds}


   \author{R.K. Prinja\inst{1}, D. Massa\inst{2}
          \and
          S.C. Searle\inst{1}
          }

   \offprints{R.K. Prinja}

   \institute{Department of Physics {\&} Astronomy, University
College London, Gower Street, London WC1E 6BT, U.K.\\
              \email{rkp@star.ucl.ac.uk,scs@star.ucl.ac.uk}
         \and
             SGT, Inc., Code 681.0, NASA Goddard Space Flight Center,
Greenbelt, MD 20771, USA\\
              \email{massa@taotaomona.gsfc.nasa.gov}
             }

   \date{Received September 15, 1996; accepted March 16, 1997}

\titlerunning{Ionization structure of B supergiant winds}
\authorrunning{Prinja, Massa, Searle}

\abstract{
We present empirically determined ionization conditions for the winds of 
106 luminous B0 to B5 stars observed by $IUE$.  The UV wind lines are 
modelled to extract products of mass-loss rates times ionization fractions 
($\mdot\,q_i(w)$, where $w = v/v_\infty$) for N\,{\sc v}, C\,{\sc iv}, 
Si\,{\sc iv}, Si\,{\sc iii}, Al\,{\sc iii} and C\,{\sc ii}.  We describe 
the general behaviour of the $\mdot\,q_i(w)$ and their ratios, 
demonstrating that the wind ionization {\em increases} with distance from 
the star, contrary to recent findings for O star winds.  Using empirical 
mass-loss rates (from H$\alpha$ observations) and model 
prescriptions, we derive mean $q_i(w)$ values integrated over the wind, 
$\langle{q_i}\rangle$.  These $\langle{q_i}\rangle$ are quite small, never 
exceeding 15\% for Al~{\sc iii} or 2{\%} for Si~{\sc iv}.  This is 
surprising, since the $\langle{q_i}\rangle$ for these ions clearly peak 
within the observed spectral range.  We conclude that the low 
$\langle{q_i}\rangle$ arise because the $\langle\mdot\,q_i\rangle$ are 
underestimated by the wind models, which assume that the outflows are 
smooth when they are, in fact, highly structured.

\keywords{stars: early-type -- stars: mass-loss -- ultraviolet: stars}
}

\maketitle
\begin{figure*}
  \begin{center}
      \includegraphics[scale=0.73]{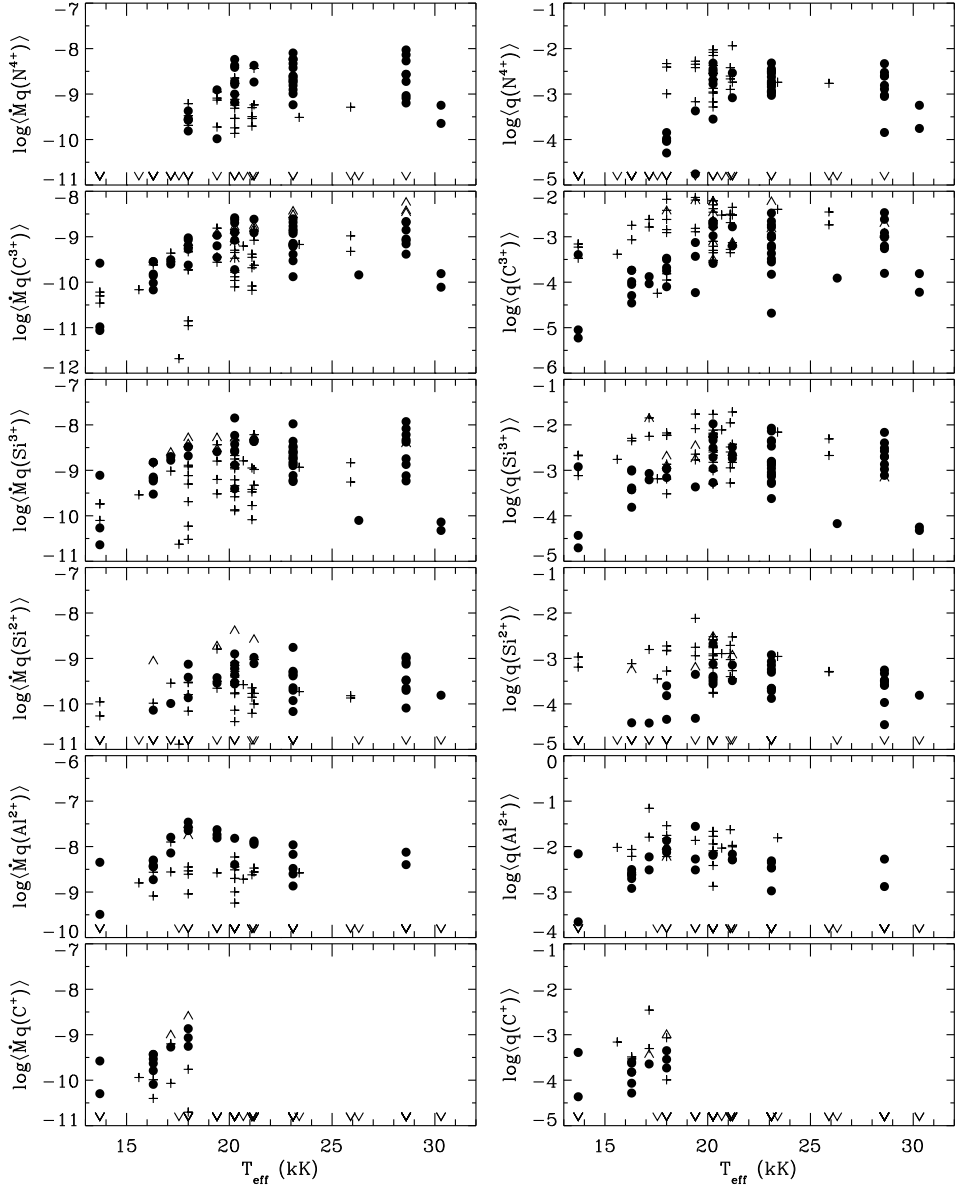}
  \end{center}
  \vspace{-.3in}
  \caption{Left panels: $\langle\mdot\,q_i\rangle$ values (integrated over 
  $0.2 \le v/v_\infty \le 0.8$) versus $T_{\rm eff}$ for the ions studied.
  Right panel: $\langle{q_i}\rangle$ versus $T_{\rm eff}$ for the same 
  ions.  Solid points are measurements from unsaturated lines of stars 
  with $\log(L/L_\odot) \geq 5$, crosses are for similar lines from stars 
  with $\log(L/L_\odot) < 5$.  Upward arrowheads indicate lower limits 
  determined from saturated lines ($\langle\tau_{rad}\rangle >$ 5 for 
  Si$^{2+}$ and C$^{2+}$, 6 for C$^{3+}$ and 7 for N$^{4+}$, Si$^{3+}$ 
  and Al$^{2+}$), and downward arrowheads indicate upper limits from 
  non-detections ($\tau_{rad}(w) \leq 0.1$ for all $w$).  }
  \label{fig1}%
\end{figure*}

\section{Introduction}
Luminous massive stars are key constituents of galaxies since they 
provide prolific ionizing radiation and an input of energy and matter
to the surrounding environment through strong radiatively driven winds.
More specifically, early-type B supergiants ($T_{\rm eff}$ $\sim$
12 to 30kK) are the most numerous hot luminous stars. They make 
substantial contributions to the integrated light of starbursts 
and provide diagnostics of star formation at low and high redshifts 
(e.g. de Mello et al. 2000).  B supergiants are also key markers in the 
determinations of extra-galactic distances based on the 
Wind-Momentum-Luminosity (WML) relation (e.g. Kudritzki {\&} Przybilla 
2003).

With this in mind, it is clearly important to understand the mass-loss
process in B supergiants and derive reliable parameters for mass-loss 
rates, $\mdot$, ionization mixture and wind structure. In this Letter we 
focus on the empirical ionization structure, as part of a wider 
comprehensive survey of B supergiants which will be published in a 
forthcoming paper (Searle et al., in preparation).  Line-synthesis 
modelling has been 
employed to derive the wind properties of a large sample of B0-B5 stars, 
for ions with resonance lines available in high-resolution $IUE$ 
data.  Our primary aim here is to examine the properties of the ionization 
state of the winds, determine whether specific ions are near the dominant 
stage of ionization, and examine how the properties change with spectral 
type and distance from the star.  Our empirical results provide 
critical constraints for more detailed wind modelling.  We highlight 
the unexpectedly low values that are apparent for the ionization 
fractions and the potential role of wind clumping.

\section{SEI modelling of $IUE$ spectroscopic data}
Our study of Galactic B supergiant winds is based on high-resolution 
($\lambda/\Delta\lambda \sim 10^4$) $IUE$ SWP 
($1150 \leq \lambda \leq 1900${\AA}) spectra of 106 stars. The individual 
spectra were binned on a wavelength grid with regular sampling of 
0.1{\AA}, and the typical signal-to-noise in the continuum is $\sim$ 20.  
The UV spectra of B supergiants contain wind lines from a wide range of 
ionization states, including 
C\,{\sc ii}\,$\lambda\lambda$1334.53, 1335.71,
C\,{\sc iv}\,$\lambda\lambda$1548.20, 1550.77,
N\,{\sc v}\,$\lambda\lambda$1238.82, 1242.80,
Al\,{\sc iii}\,$\lambda\lambda$1854.72, 1862.79,
Si\,{\sc iii}\,$\lambda\lambda$1206.50 and
Si\,{\sc iv}\,$\lambda\lambda$1393.76, 1402.77.

To extract physical parameters from the wind profiles, we used the 
methods described in detail by Massa et al.\ (2003).  They employ a 
modified version of the `Sobolev with exact integration' (SEI) code 
(Lamers et al.\ 1987).  Once the velocity field of the wind is determined, 
this approach provides reliable radial optical depths for unsaturated wind 
lines as a function of normalized velocity, $\tau_{rad}(w)$, where 
$w = v/v_\infty$ and $v_\infty$ is the terminal velocity of the wind.  The 
fits to individual wind lines require parameters for the terminal 
velocity, `$\beta$-type' velocity law, turbulent velocity, and 
$\tau_{rad}(w)$ (specified as a set of 21 independent velocity bins that 
are adjusted by a non-linear least squares procedure). Photospheric 
spectra were supplied using the $IUE$ line profiles of the non-supergiant 
B stars listed by Prinja et al.\ (2002).  The $\tau_{rad}(w)$ were 
converted into a product of $\mdot$ times ionization fraction, 
$\mdot\,q_i(w)$ (see, e.g., Massa et al.\ 2003).  To isolate $q_i(w)$ for 
each ion requires an estimate of $\mdot$.  We discuss our approach to this 
problem in \S~3.1.  

\section{Ionization conditions}
Figure~1 (left panels) shows mean $\langle\mdot\,q_i\rangle$ values over 
0.2 $\le$ $v/v_\infty$ $\le$ 0.8, for the ions analysed versus 
$T_{\rm eff}$, and (right panels) the mean $\langle{q_i}\rangle$ values 
for the same ions (see \S~3.1).  We adopted the Humphreys {\&} McElroy 
(1984) spectral type -- $T_{\rm eff}$ calibration.  Recent detailed NLTE 
model atmosphere analyses (e.g. Crowther et al.\ 2002, Bianchi {\&} Garcia 
2002, Herrero et al.\ 2002) have revised the $T_{\rm eff}$ scale for 
O-stars downward by $\sim$ 10 $-$ 20{\%}.  There are also indications that 
B star temperatures should be reduced as well (e.g. Evans et al.\ 2004, 
Searle et al.\ 2005). However, while the revised $T_{\rm eff}$ of a B0 Ia 
may be $\sim 2000$~K lower than those used in Fig. 1, the differences are 
expected to be $\la 5${\%} at B1 and later (Searle et al.\ 2005).

We first consider the $\langle\mdot\,q_i\rangle$.  These behave as 
expected: at fixed $T_{\rm eff}$, all of the $\langle\mdot\,q_i\rangle$ 
tend to be larger for more luminous stars, with some overlap, 
and the presence of an ion at a given $T_{\rm eff}$ is related to the 
energies required to produce and destroy it.  Specifically, 
$\langle\mdot\,q($N$^{4+})\rangle$ and $\langle\mdot\,q($C$^{3+})\rangle$ 
increase over the entire $T_{\rm eff}$ range.  $\langle\mdot\,q($Si$^{3+})
\rangle$ appears to peak (or at least plateau) at $\sim 20$kK (with 
considerable spread at each $T_{\rm eff}$).  The peak is more distinct in 
the less luminous stars, where saturation is not a problem.  $\langle
\mdot\,q($Si$^{2+})\rangle$ (which can be measured in fewer stars) peaks 
at $\sim$ 20kK.  $\langle\mdot\,q($Al$^{2+})\rangle$ peaks at 18kK with 
a value $\ga 3\times 10^{-8}$ M$_\odot$ yr$^{-1}$, the largest of all ions.
Al\,{\sc iii} is a useful ion for mass-loss studies of cooler B stars, 
since it rarely saturates (due to its relatively low abundance) and is 
unaffected by nuclear processing.  Note that $\langle\mdot\,q($Al$^{2+})
\rangle$ peaks at a lower $T_{\rm eff}$ than $\langle \mdot\,q($Si$^{2+})
\rangle$, even though a higher energy is required to produce it.  This 
occurs because it is destroyed at a lower energy than Si$^{2+}$.  $\langle
\mdot\,q($C$^{+})\rangle$ peaks at 18kK and then abruptly drops to zero -- 
independently of the stellar luminosity.  This unique behaviour may be due 
to the near coincidence of the ionization potentials of C$^{+}$ (24.38~eV) 
and He (24.48~eV).  
When He ionizes in the stellar photosphere, the stellar radiation field in 
the He$^{+}$ continuum ($228$ $\leq$ $\lambda$ $\leq$ $506$~\AA) increases 
dramatically.  Indeed, Kurucz (1991) $\log g=3$ models with $T_{\rm eff}=
17$ and 19~kK show that the integrated flux in this region increases by a 
factor of 30.  This enormous increase could convert all of the C$^{+}$ in 
the wind to C$^{2+}$. However, these conjectures must be verified by detailed 
modelling.

   \begin{figure}
      \includegraphics[scale=0.7]{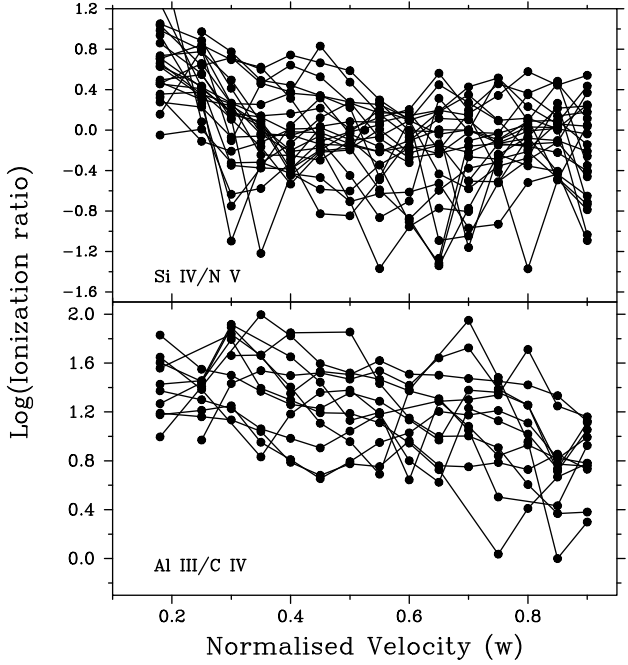}
      \caption{Ionization fraction ratios of
Si$^{3+}$/N$^{4+}$ (upper panel) and Al$^{2+}$/C$^{3+}$
(lower panel) plotted as a function of normalised velocity
($w$ = $v/v_\infty$). Only the cases of unsaturated line profiles
are shown.
              }
         \label{fig2}%
   \end{figure}
%


We now consider $\mdot\,q$ ratios, which are independent of $\mdot$.  We 
begin with ratios of the $\langle\mdot\,q_i\rangle$, which provide insight 
into the how the wind ionization responds to $T_{\rm eff}$.  The overall 
trend is that the empirical ionization ratios of N$^{4+}$/Si$^{3+}$, 
C$^{3+}$/Si$^{3+}$ and Si$^{3+}$/Al$^{2+}$ increase as a function of 
$T_{\rm eff}$, indicating that the winds of hotter B supergiants are more 
highly ionized.  Some of the ratios straddle the `bistability jump' at 
$\sim$21kK, where there is a ramped increase in the ratio of terminal 
velocity to escape velocity (e.g. Lamers et al.1995, Vink et al. 2000). 
However, the ratios of C$^{3+}$/Si$^{3+}$, and Si$^{3+}$/Si$^{2+}$ do not 
suggest a dramatic change in the ionization of the winds across the 
`jump', though the former has a steeper drop below $\sim$ 20kK.  

   \begin{figure}
      \includegraphics[scale=0.5]{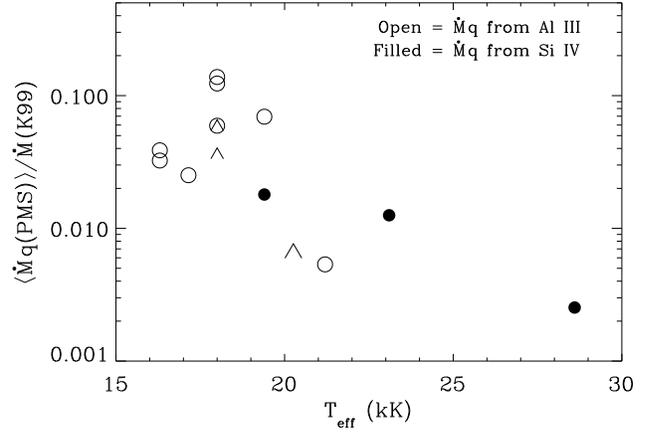}
      \caption{Mean ionization fraction of
Si$^{3+}$ (filled) and Al$^{2+}$ (open), based on the H$\alpha$
mass-loss rates of Kudrtitzi et al. (1999; K99).
              }
         \label{fig3}%
   \end{figure}
%


Finally, we examine how the ionization of B supergiant winds varies with 
distance from the star by plotting $\mdot\,q_i(w)$ ratios for 
Si$^{3+}$/N$^{4+}$ and Al$^{2+}$/C$^{3+}$ against $w$ ($=v/v_\infty$) in 
Fig.~2 (for unsaturated cases only).  Broadly, the ratios decrease over 
$0.2$ $\leq$ $w$ $\leq$ $0.9$ (typically $1.2 \leq r/R_\star \leq 10$), 
implying an {\it increase} in the ionization state with velocity. This is 
the opposite of the trend found by Massa et al.\ (2003) in LMC O stars, 
and is more in agreement with the expectations of the optically thin 
nebular approximation (e.g., Cassinelli \& Olson 1979).  It could indicate 
a fundamental difference between the winds of O and B-type stars.  
We intend to compare these trends to detailed model predictions in a future 
publication.

\subsection{Mean ion fractions and $\mdot$}

Conversion of our $\langle\mdot\,q_i\rangle$ measurements into 
$\langle q_i\rangle$, requires $\mdot$ values.  We consider two options.  
First we adopt empirical mass-loss rates determined from a NLTE model 
atmosphere analysis of H$\alpha$ profiles by Kudritzki et al.\ (1999).  
This accounts for only 14 stars in our sample, but it allows us to compare 
our results to previous work.  The resulting ion fractions for Al\,{\sc 
iii} and Si\,{\sc iv} are shown in Fig.\ 3, where we have adjusted the 
Kudritzki et al. $\mdot$s so that the adopted radii and temperatures are 
consistent with those used in Fig.~1 (the adjustments are typically less 
than a factor of 2, and not systematic).
We see that $\langle q($Al$^{2+}) \rangle$ peaks at 
$\sim$ 10--15{\%}, which is comparable to the maximum value of 
$\langle q($P$^{4+}) \rangle$ found by Massa et al.\ (2003, 2004) for O 
stars mass-loss rates based on the far-UV P\,{\sc v} wind lines and either 
empirical or theoretical $\mdot$s.  The $\langle q($Si$^{3+}) \rangle$ 
peak is poorly sampled by the data and dominated by low luminosity stars, 
since the more luminous stars have saturated Si~{\sc iv} wind lines.  

Second, since the vast majority of our program stars do not have 
empirically determined mass-loss rates, we use the Vink et al.\ 
(2000, 2001) model prescriptions between stellar parameters and $\mdot$.
Note that changes of $T_{\rm eff}$ or $L/L_\odot$ resulting from a $\pm$ 
one spectral or luminosity bin error, change these predictions by less than 
a factor of 2, and that these values are far larger than differences 
between our adopted parameters and the ones determined by Kudritzki et al.\ 
(1999) for stars in common. The right side of Fig.~1 shows the resulting 
$\langle{q_i}\rangle$ as a function of $T_{\rm eff}$.  Overall, these 
follow the same trends as the $\langle\mdot\,q_i\rangle$.  While it {\em 
appears} that the {\em less} luminous stars have larger $\langle{q_i}
\rangle$, this is most likely a result of line saturation in the more 
luminous stars.

We now consider the four ions which peak in the B star range, to 
determine if any of them are dominant.  We find that $\langle 
q($Si$^{3+})\rangle$ never exceeds 1.9\%, and that the mean of all 
unsaturated cases near its peak is $0.48$\%.  These quantities for the 
other ions are: 1.2 and 0.14\% (Si$^{2+}$), 6.9 and 1.7\% (Al$^{2+}$) 
and 0.34 and 0.08\% (C$^{+}$).  The $\langle q($Al$^{2+}) \rangle$ values 
are based primarily on high luminosity stars, and these exceeds 2.5\% in 
only one case -- somewhat smaller than Massa et al.\ found for P~{\sc v} 
in the O stars. The $\langle q($Si$^{3+}) \rangle$ values are particularly 
surprising, since models predict it should be $\sim 1$ for $18$ $\leq$ 
$T_{\rm eff}$ $\leq$ $20$kK.  However, due to the saturation of the 
Si~{\sc iv} resonance lines in the more luminous stars, the Si~{\sc iv} 
result is based exclusively on the less luminous stars in our sample, and 
Vink et al. (2000) have already pointed out that there is a large 
discrepancy between their predicted B star mass-loss rates and values 
based on H$\alpha$ line profiles for stars with log($L/L_\odot$)$\la$ 
5.8.

\section{Implications for the nature of the mass-loss}

The major conclusions of the Letter are:
\begin{enumerate}
\item The ionization structure of the winds of B supergiants is different 
from the O stars (Fig.\ 2).  We caution, however, that most of these 
results are for low luminosity stars.

\item There is no evidence of an abrupt change of the wind ionization in 
the temperature region of the bistability jump.

\item The ionization fractions of non-CNO ions never exceed 0.15 (for 
Al~{\sc iii}) in the B supergiants -- a result consistent with P~{\sc v} 
based results in the O stars (Massa et al.\ 2004).

\item The ionization fractions of Si~{\sc iv} never exceeds 0.02. 
This result is based exclusively on lower luminosity B supergiants (where 
the Si~{\sc iv} lines are unsaturated) and holds when using either theoretical 
or empirical mass loss rates.  It suggests that the theory is 
incomplete for lower luminosity stars and underscores the need for more 
empirical mass loss rates for these stars.
\end{enumerate}

For Galactic B supergiants, abundance anomalies cannot explain the 
very low $\langle{q_i}\rangle$ for Al and Si, since these elements have 
well known abundances and are unaffected by stellar evolution.
Revisions to the B star $T_{\rm eff}$ scale are expected to have little 
affect on the predicted mass-loss rates at the temperatures where 
Si$^{3+}$, Si$^{2+}$ and Al$^{2+}$ peak (see \S~3).  Note also that for 
this key temperature range, stellar radii based on the Humphreys {\&} 
McElroy (1984) calibration differ from the Kurdtizki et al.\ (1999) 
radii by only $\sim 5 - 15${\%}.
For the range of temperatures and ionizations considered in this Letter, 
it is highly unlikely that none of the ions are remotely close to 
dominant.  As Massa et al.\ (2003) discuss concerning a similar result 
for P\,{\sc v} in LMC O-type stars, we suspect that the small ion 
fractions indicate clumping in B supergiant winds.  UV time-series 
spectroscopy has clearly demonstrated that the winds of early-B 
supergiants are highly variable and structured, with evidence for 
rotational modulation in many cases (e.g. Prinja et al. 2002), and optical 
studies (see, Kaufer 1999) have reached the same conclusions for 
mid-to-late B supergiants.  Furthermore, Blomme et al.\ (2002) find 
excess flux at millimetre wavelengths in $\epsilon$ Ori (B0 Ia) compared 
to homogeneous wind model predictions, indicating considerable wind 
structure at several tens of stellar radii from the photosphere.  
Note, that if the winds are extremely clumped or structured, we must also 
re-evaluate the mass loss 
rates inferred from the radio and H$\alpha$ observations.  As a 
result, it is difficult at this time to infer how much of the low 
ionization fractions result from systematics in the measurement 
processes, and how much results from overestimates of the mass loss 
rates.

\begin{acknowledgements}
DM acknowledges support from NASA contract ADP03-0031-0123.  SCS thanks 
PPARC for financial support. We are grateful for the comments and 
suggestions of the referee.

\end{acknowledgements}

{}


\end{document}